\documentclass[12pt]{iopart}

\usepackage{graphicx}% Include figure files
\usepackage{dcolumn}% Align table columns on decimal point

\newcommand{\vect}[1]{{\bf #1}}

\newcommand{\beq}{\begin{equation}}
\newcommand{\eeq}{\end{equation}}
\newcommand{\beqar}{\begin{eqnarray}}
\newcommand{\eeqar}{\end{eqnarray}}

\begin{document}
\title{Playing Games with the Quantum Three-Body Problem }
\author{Tarun Biswas}
\address{State University of New York at New Paltz, \\ New Paltz,  NY 12561, USA.}
\ead{biswast@newpaltz.edu}

\begin{abstract}
Quantum mechanics courses focus mostly on its computational aspects.
This alone does not provide the same depth of understanding as most physicists
have of classical mechanics. The understanding of classical mechanics is
significantly bolstered by the intuitive understanding that one acquires
through the playing of games like baseball at an early age. It would be
good to have similar games for quantum mechanics. However, real games that
involve quantum phenomena directly are impossible. So, computer simulated
games are good alternatives.
Here a computer game involving three interacting quantum particles
is discussed. It is hoped that such games played at an early age will provide
the intuitive background for a better understanding of quantum mechanics
at a later age.
\end{abstract}

\section{Introduction}
``\ldots and then the wavefunction collapses.'' What visual images are
inspired by such a statement? We can visualize collapsing bridges, buildings
and maybe even a souffl\'{e}. But collapsing wavefunctions are a visual mystery
for both novices and experts in quantum mechanics. A few years of graduate
school can teach a physics student the mathematical methods as well as the 
experimental tests of quantum mechanics. However, acquiring an intuitive
understanding of the subject is more challenging. Classical mechanics is easier 
to understand due to the ready availability of visual images (collapsing
bridges, souffl\'{e}s, etc.). It is also significant that these images of
classical mechanics are observed by everyone at an early age making them
part of our intuition. Similar early introductions to quantum phenomena would
be very useful for the learning experience of children. They would build a
foundation for later, more rigorously mathematical, presentations of quantum
mechnics. However, real visual images for quantum mechanics are difficult 
to find. So let us look for some computer simulated visual images through a 
computer game based on quantum mechanics.

Using computers for physics education has become quite mainstream through the last 
decade\cite{cook,ehrlich,christ,kansas}. However, using physics based computer games 
is not that common\cite{bis1,bis2}. Personally, I prefer the game
approach for several reasons. It can provide an intuitive understanding at
a very early age without the need for mathematics. It is non-threatening and develops physics
intuition in a relaxed setting. It is like understanding projectile motion
while playing baseball. In particular, for quantum mechanics, computer games
are uniquely useful as real games like baseball shed very little light on
the subject.

In the past, I have developed a game based on the quantum mechanical free
particle (``Quantum Duck Hunt'')\cite{bis3,qduck}. Here, I present a
significantly more complex system -- an interacting three particle
system. The game based on this system is called ``Quantum Focus''\cite{qfocus}.
It deals with various subtle aspects of quantum observation and wavefunction
collapse\cite{bis4}.

This game is not meant to teach quantum mechanics to college students. It is meant
to develop ``quantum intuition'' at a much earlier age (maybe elementary school). 
Children playing this kind of games are expected to develop ``gut feelings'' about
quantum phenomena just as they usually do for classical
phenomena by playing baseball or soccer. The approach here is not that of standard 
accepted padagogy. Something different is being tried to see if 
it works better. I have tried it on a few children (my own and their 
friends!). The results are very encouraging -- now they want to learn
quantum mechanics! If we wait until college to develop quantum
intuition in youngsters, we might miss the formative years when most
intuition is developed. However, college students may also benefit from this game by
studying the computer code and trying to come up with their own variations.

\section{The game}
\label{secgame}

\begin{figure}
\includegraphics[scale=0.7]{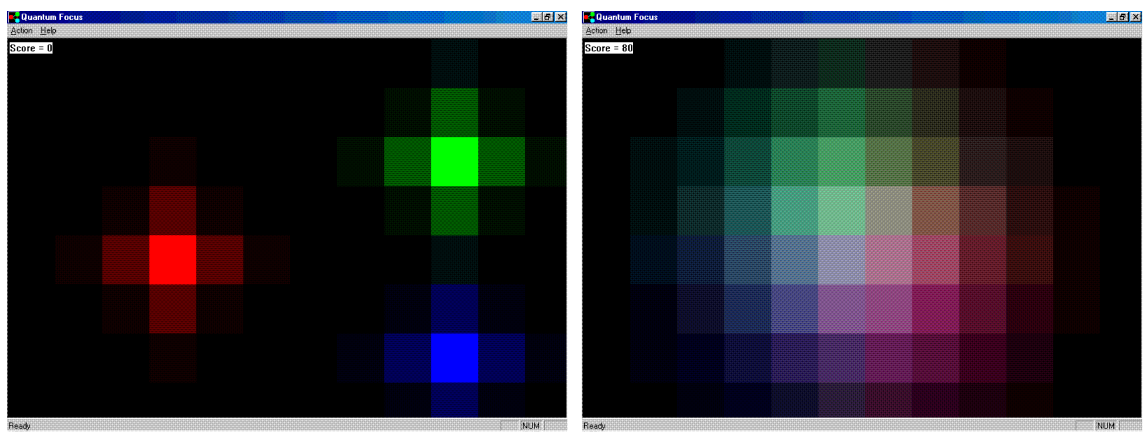}
\caption{The game screen at starting time (left) and at a later time (right).
 \label{fig1}}
\end{figure}

The game is started by choosing ``Start'' from the ``Action'' menu. Three boxes
colored red, green and blue appear on a black background screen. With time, the color
of each box begins to smear into neighboring boxes (see figure~\ref{fig1}). 
At the same time, the brightest spot in each color smear moves away from the other two.
As the colors spread, they produce mixtures of the primary colors in various
proportions in different boxes on the screen.

The object of the game is to make each color smear as small as possible and at the same
time bring the three colors as close together as possible. The only tool available
for achieving this is the click of the mouse at strategic points. Clicking the mouse
will retract a color completely into a single box -- the one that was clicked. 
But this retraction or ``collapse'' will occur only with a probability proportional
to the preexisting intensity of that color in that box. So, if there is very little
red in a certain box, it is unlikely that red can be collapsed into it by any
amount of clicking. This is why, sometimes, a click of the mouse may produce
no effect at all. An interesting sound effect will accompany the actual collapse
of a color.

The quantum mechanically minded readers must already have noted that the three
primary colors represent three particles. The intensity of a primary color is
related to the probability of finding the corresponding particle at a given place.
The mouse clicks simulate a particle detector's attempts at detecting a particle.
The retraction of a color into a single box simulates a wavefunction collapse.
In the present model, the three particles are tied together by attractive forces. 
So, it is a bit tricky to see why quantum mechanics makes the three color smears 
move apart. This effect will be discussed later.

A score is computed in each time step. It depends on how
small each color smear is and how close the three colors are. So, the goal is to 
produce a single white box and no other colored boxes. But this state of perfection
can be seen to be impossible. The score displayed (at the top left corner) is 
the maximum score achieved during the course of a game.

It should be noted that, while the colors spread, nothing is lost. Colors that
spread off-screen on one edge reappear on the opposite edge. This effect may be
used for game strategies.

The game can be played at four levels of difficulty. The scoring formula respects
the level of difficulty. The features of these levels will be discussed later.

\section{The quantum three particle problem}
This game is based on the quantum dynamics of three interacting distinguishable
particles. Most quantum problems deal with the solution of the time-independent
Schr\"{o}dinger equation. But here we are concerned with the wavefunction
collapse and subsequent change in the wavefunction. Hence, it is necessary to
solve the time-dependent Schr\"{o}dinger equation\cite{bis4}.

Let the positions of the three particles be $\vect{q}_{i}$, and their
momenta be $\vect{p}_{i}$ ($i=1,2,3.$). Let the wavefunction of the system
be $\psi(\{\vect{q}_{i}\},t)$ and its hamiltonian 
$H(\{\vect{q}_{i}\},\{\vect{p}_{i}\})$ where $t$ is time. Then the
time-dependent Schr\"{o}dinger equation to be solved is\cite{shankar,bis5}
\beq
i\hbar\frac{\partial\psi}{\partial t}=H\psi \label{eq1}
\eeq
\begin{figure}
\hfil\includegraphics{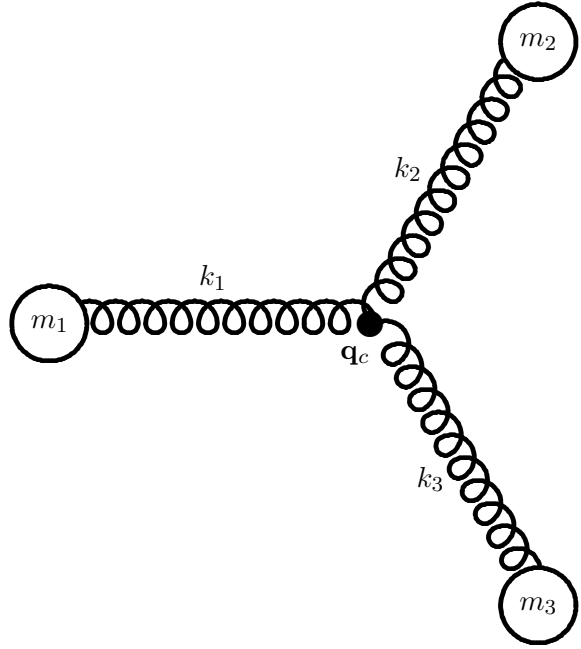}\hfil
\caption{Three particles attached by ``springs'' \label{fig2}}
\end{figure}

For simplicity I choose harmonic (``spring'') potentials to represent the
interparticle forces (see figure~\ref{fig2}). Also, The springs are assumed 
to have zero unextended length. The resulting hamiltonian is as follows.
\beq
H=\sum_{i=1}^{3}\frac{\vect{p}_{i}^{2}}{2m_{i}} + 
\sum_{i=1}^{3}\frac{k_{i}}{2}(\vect{q}_{i}-\vect{q}_{c})^{2}, \label{eq2}
\eeq
where $m_{i}$ is the mass and $k_{i}$ the spring constant for the $i^{\mbox{th}}$
particle. $\vect{q}_{c}$ is the position of the common center at which the three
springs are tied.

It can be seen that, like most three particle problems, this cannot be separated
in variables. The dependence on $\vect{q}_{c}$ complicates the hamiltonian
significantly as it is not an independent coordinate. $\vect{q}_{c}$ depends on 
the particle coordinates due to the following zero net force condition.
\beq
\sum_{i=1}^{3}k_{i}(\vect{q}_{i}-\vect{q}_{c})=0, \label{eq3}
\eeq
which gives
\beq
\vect{q}_{c}=\frac{\sum_{j=1}^{3}k_{j}\vect{q}_{j}}{\sum_{j=1}^{3}k_{j}}. \label{eq4}
\eeq
Hence,
\beq
H=\sum_{i=1}^{3}\frac{\vect{p}_{i}^{2}}{2m_{i}} + 
\sum_{i=1}^{3}\frac{k_{i}}{2}
\left(\vect{q}_{i}-\frac{\sum_{j=1}^{3}k_{j}\vect{q}_{j}}{K}\right)^{2}, \label{eq5}
\eeq
where $K=\sum_{j=1}^{3}k_{j}$.

Equation~\ref{eq1} can be solved numerically to obtain the time development
of the wavefunction provided an intial value is specified. At any point in this
time development, if a particle detector detects the first particle 
in a small region $R$, it will be with a probability
\beq
P_{1} = \int_{R}d^{3}\vect{q}_{1}\int_{-\infty}^{\infty}d^{3}\vect{q}_{2}
\int_{-\infty}^{\infty}d^{3}\vect{q}_{3}
\psi^{*}(\{\vect{q}_{i}\},t)\psi(\{\vect{q}_{i}\},t). \label{eq6}
\eeq
If the particle is actually detected, the wavefunction must collapse to
\beq
\psi_{c}(\{\vect{q}_{i}\},t)=A\Delta_{R}(\vect{q}_{1}-\vect{q}_{0})
                         \psi(\{\vect{q}_{i}\},t), \label{eq7}
\eeq
where $\Delta_{R}(\vect{q}_{1}-\vect{q}_{0})$ is a sharply peaked
function that is nonzero only in the region of detection $R$ centered
about the position $\vect{q}_{0}$. The detailed form of this function
depends on the detector sensitivity in the region $R$. In the limit
$R\rightarrow 0$, it is the Dirac delta function:
\beq
\lim_{R\rightarrow 0}\Delta_{R}(\vect{q}_{k}-\vect{q}_{0})
      =\delta^{3}(\vect{q}_{k}-\vect{q}_{0}). \label{eq8}
\eeq
The constant $A$ is needed to renormalize $\psi$ after the collapse.
The detection of the other two particles can be described similarly.

After the collapse, $\psi$ is replaced by $\psi_{c}$ and the time development
continued as given by equation~\ref{eq1} until the next collapse.

\section{The numerical technique}
The computer screen being 2-dimensional, the above formulation will be reduced
to its 2-dimensional equivalent for the purpose of the game. To use standard 
finite difference methods, the screen
space is divided into a matrix of $m$ columns and $n$ rows to produce a total of
$m\times n$ boxes. For the purpose of a game, we may sacrifice accuracy for speed
as long as the qualitative aspects of the system are maintained. So, the maximum
values of $m$ and $n$ are chosen to be 12 and 8.

To solve the Schr\"{o}dinger equation, boundary conditions must be specified. There
are several possible natural choices:
\begin{enumerate}
\item Perfectly reflecting boundary conditions.
\item Perfectly absorbing boundary conditions.
\item Periodic boundary conditions.
\end{enumerate}
The perfectly reflecting
boundary produces a discontinuity at the boundary that interferes with visualization.
The perfectly absorbing boundary allows particles to go off screen, thus making
them useless for visualization. The periodic condition seems to be the best for 
visualization. It identifies the left edge to the right and the bottom edge to the
top (toroidal topology). Hence, particle current that disappears on one edge reappears
on the opposite edge.

The discrete forms for the $x$ and $y$ components of each coordinate $\vect{q}_{i}$
may be written as
\beq
q_{ix}=a_{ix}\Delta x,\;\;\;\;  q_{iy}=a_{iy}\Delta y, \label{eq9}
\eeq
where $i=1,2,3$, $a_{ix}=0,1,2,\ldots ,m-1$, and $a_{iy}=0,1,2,\ldots ,n-1$.
$\Delta x$ is the mesh width in the $x$ direction and $\Delta y$ is the mesh width
in the $y$ direction. If the mesh width in time is $\Delta t$, then 
equation~\ref{eq1} produces the following recursive formula for the computation
of $\psi$.
\beq
\psi(\{\vect{q}_{i}\},t) = (1-iH\Delta t/\hbar)\psi(\{\vect{q}_{i}\},t-\Delta t). \label{eq10}
\eeq
In general, the above numerical algorithm for the solution of the Schr\"{o}dinger
equation is known to be unstable\cite{Teukolsky}. However, we can use it in the present case because wavefunction collapses are expected to preempt any instability. Besides,
as noted before we are not looking for high accuracy.

The wavefunction $\psi$, at one instant of time, is a function of all coordinates
$\vect{q}_{i}$. So, its discretized form must depend on all $a_{ix}$ and $a_{iy}$.
Thus, for numerical computation, $\psi$ is represented by an array of 6 dimensions
(one for each $a_{ix}$ and $a_{iy}$). In the notation of the C language it
would be: $\psi[a_{1x}][a_{1y}][a_{2x}][a_{2y}][a_{3x}][a_{3y}]$. 
For compactness of notation I can write this as:
$\psi[a][b][c][d][e][f]$ or $\psi_{a,b,c,d,e,f}$. Then the finite 
difference form of the operation
by the hamiltonian $H$ is found from equation~\ref{eq5} using equation~\ref{eq9}
and the following finite difference forms of the $\vect{p}_{i}^{2}$ operators.
\beqar
\vect{p}_{1}^{2}\psi_{a,b,c,d,e,f} & = & -\hbar^{2}\left(
\frac{\psi_{a+1,b,c,d,e,f}-2\psi_{a,b,c,d,e,f}+\psi_{a-1,b,c,d,e,f}}
{(\Delta x)^{2}}+\right. \nonumber \\*
&&\left.+\frac{\psi_{a,b+1,c,d,e,f}-2\psi_{a,b,c,d,e,f}+\psi_{a,b-1,c,d,e,f}}
{(\Delta y)^{2}}\right), \nonumber \\
\vect{p}_{2}^{2}\psi_{a,b,c,d,e,f} & = & -\hbar^{2}\left(
\frac{\psi_{a,b,c+1,d,e,f}-2\psi_{a,b,c,d,e,f}+\psi_{a,b,c-1,d,e,f}}
{(\Delta x)^{2}}+\right. \nonumber \\*
&&\left.+\frac{\psi_{a,b,c,d+1,e,f}-2\psi_{a,b,c,d,e,f}+\psi_{a,b,c,d-1,e,f}}
{(\Delta y)^{2}}\right), \nonumber \\
\vect{p}_{3}^{2}\psi_{a,b,c,d,e,f} & = & -\hbar^{2}\left(
\frac{\psi_{a,b,c,d,e+1,f}-2\psi_{a,b,c,d,e,f}+\psi_{a,b,c,d,e-1,f}}
{(\Delta x)^{2}}+\right. \nonumber \\*
&&\left.+\frac{\psi_{a,b,c,d,e,f+1}-2\psi_{a,b,c,d,e,f}+\psi_{a,b,c,d,e,f-1}}
{(\Delta y)^{2}}\right).  \label{eq11}
\eeqar
Here the most common finite difference form for second derivatives is used.
Using equations~\ref{eq5}, \ref{eq9}, and \ref{eq11} in equation~\ref{eq10}, 
the wavefunction for
successive time steps can be computed. The numerical method chosen here
does not maintain normalization of $\psi$. Hence, after
each time step computation, $\psi$ must be normalized\cite{Teukolsky}.

Also after each time step computation, the screen image must be updated to
provide an animated visual effect. The RGB coloring scheme on the computer
screen provides a natural way of representing the three particle probabilities.
The red, green and blue color intensities in a box are made proportional
to the probabilities of finding each of the three particles in that box. The
discretized version of equation~\ref{eq6} provides the probabilities to be
used. They are as follows.
\beqar
P_{1} & = & \sum_{c,d,e,f}\psi^{*}_{a,b,c,d,e,f}\psi_{a,b,c,d,e,f}, \nonumber \\
P_{2} & = & \sum_{a,b,e,f}\psi^{*}_{a,b,c,d,e,f}\psi_{a,b,c,d,e,f}, \nonumber \\
P_{3} & = & \sum_{a,b,c,d}\psi^{*}_{a,b,c,d,e,f}\psi_{a,b,c,d,e,f}. \label{eq12}
\eeqar
So, the C++ code used to find the amount of red color (particle 1) in a box is as follows.
\begin{verbatim}
int CQFocusDoc::Red(int a, int b)
{
      int c, d, e, f;
      float red = 0;

      for(c=0;c<xpts;c++)
            for(d=0;d<ypts;d++)
                  for(e=0;e<xpts;e++)
                        for(f=0;f<ypts;f++)
                              red += norm(Psi[a][b][c][d][e][f]);

      return(255*sqrt(sqrt(red))); //Square root used to enhance
                                   //color for better visibility.
}
\end{verbatim}
The amounts of the other two colors are computed similarly.

When the mouse button is clicked in a box, one of the three particles is picked
randomly for collapse and then the decision to actually collapse it is made
based on the probability given by equation~\ref{eq12}. The following C++ code
fragment makes these probabilistic decisions.
\begin{verbatim}
partnum = MyRandom(3); //Generates an integer random number between 0 and 2.
intensity = MyRandom(256);
intensity = (intensity*intensity*intensity)/(256*256);  
// The above redifinition improves game by requiring less mouse clicking.

i = point.x/bwidth; // x pixel position divided by box width.
j = point.y/bheight; // y pixel position divided by box height.

switch(partnum)
{
case 0:
  if(intensity < pDoc->Red(i,j)) // Function Red(i,j) defined above.
    { pDoc->Collapse(partnum, CPoint(i,j)); setcollapse = true;}
  break;
case 1:
  if(intensity < pDoc->Green(i,j))
    { pDoc->Collapse(partnum, CPoint(i,j)); setcollapse = true;}
  break;
case 2:
  if(intensity < pDoc->Blue(i,j))
    { pDoc->Collapse(partnum, CPoint(i,j)); setcollapse = true;}
  break;
}
\end{verbatim}
The wavefunction after the collapse is given by equation~\ref{eq7}. The function 
$\Delta_{R}(\vect{q}_{k}-\vect{q}_{0})$ in its discrete form is chosen as
the discrete form of the Dirac delta function:
\beq
\Delta_{R}(\vect{q}_{k}-\vect{q}_{0})=\left\{\begin{array}{ll}
                                        1, & \mbox{if $a_{kx}=a_{0x}$ and $a_{ky}=a_{0y}$}\\
                                        0, & \mbox{otherwise},
                                             \end{array}
                                              \right. \label{eq13}
\eeq
where the integer values $a_{kx}$, $a_{0x}$, $a_{ky}$ and $a_{0y}$ are 
defined as in equation~\ref{eq9}. So, the C++ code for wavefunction collapse is as 
follows.
\begin{verbatim}
void CQFocusDoc::Collapse(int partnum, CPoint point)
{
  int a, b, c, d, e, f;
  switch(partnum)
  {
  case 0:
    for(a=0;a<xpts; a++)
      for(b=0;b<ypts;b++)
        for(c=0;c<xpts;c++)
          for(d=0;d<ypts;d++)
            for(e=0;e<xpts;e++)
              for(f=0;f<ypts;f++)
              {
                if(a!=point.x || b!=point.y)
                  Psi[a][b][c][d][e][f] = 0;
              }
    break;
  case 1:
    for(a=0;a<xpts; a++)
      for(b=0;b<ypts;b++)
        for(c=0;c<xpts;c++)
          for(d=0;d<ypts;d++)
            for(e=0;e<xpts;e++)
              for(f=0;f<ypts;f++)
              {
                if(c!=point.x || d!=point.y)
                  Psi[a][b][c][d][e][f] = 0;
              }
    break;
  case 2:
    for(a=0;a<xpts; a++)
      for(b=0;b<ypts;b++)
        for(c=0;c<xpts;c++)
          for(d=0;d<ypts;d++)
            for(e=0;e<xpts;e++)
              for(f=0;f<ypts;f++)
              {
                if(e!=point.x || f!=point.y)
                  Psi[a][b][c][d][e][f] = 0;
              }
    break;
  }
  Normalize();
}
\end{verbatim}
Here \verb+partnum+ gives the randomly picked particle number and
\verb+point+ identifies the box clicked.

\section{The levels of difficulty}
The game can be played at four different levels of difficulty. The difficulty
level is increased by increasing the values of the spring constants $k_{i}$.
This increases the rate at which the positions of maximum probability move
apart. The reason will be seen in the next section. 

The difficulty level is increased also by reducing the total number of boxes. 
This increases the speed of computation and hence increases the rate of
spreading of the wavefunction.

The score allows for higher difficulty levels.

\section{Some results}
The primary purpose of this game is to provide repeated and consistent
visual effects that mimic quantum wavefunction dynamics. 
As expected, the wavefunction collapse leaves the undetected particles unaffected.
Also as expected, the probability profile of each particle spreads with
time. The resulting mix of the primary colors produces some rather 
unusual color effects that may interest the artists amongst us. What is not-so-obvious
is as follows. If we start with one particle in a collapsed state (with
no velocity), with time its probability
peak moves away from those of the other particles! As the potential function used here is
attractive, this is somewhat surprising. However, closer scrutiny can explain
this phenomenon.

Consider the standard one-particle harmonic oscillator. Higher energy eigenstates
have probability peaks farther away from the origin. This means that particles that
start off with higher momenta are likely to have their probability peaks farther away.
For the present case, a particle collapsed to its position eigenstate 
has high probabilities for large momenta and hence, large energy. This makes its
probability peak move away from the other particles.

\end{document}